\documentclass[epj]{svjour}
\usepackage[dvips]{graphicx}
\usepackage{epsbox,color}
\usepackage{amsmath}
\usepackage{float}

\begin{document}
\title{A Bayesian analysis of the nucleon QCD sum rules}
\author{Keisuke Ohtani\inst{1}\thanks{Electronic address: ohtani.k@th.phys.titech.ac.jp}, 
Philipp Gubler\inst{1} \and Makoto Oka\inst{1}% etc
}                     % Do not remove
%         % Insert a name or remove this line
%
\institute{Department of Physics, H-27, Tokyo Institute of Technology, Meguro, Tokyo 152-8551, Japan}
\date{Received: date / Revised version: date}
\abstract{
QCD sum rules of the nucleon channel are reanalyzed, using the maximum entropy method (MEM). 
This new approach, based on the Bayesian probability theory, 
does not restrict the spectral function to the usual ``pole + continuum"-form, allowing 
a more flexible investigation of the nucleon spectral function. 
Making use of this flexibility, we are able to investigate the spectral functions of 
various interpolating fields, finding that the nucleon ground state mainly 
couples to an operator containing a scalar diquark. 
Moreover, we formulate the Gaussian sum rule for the nucleon channel and find 
that it is more suitable for the MEM analysis to extract 
the nucleon pole in the region of its experimental value, while the Borel sum rule does 
not contain  enough information to clearly separate the nucleon pole from the 
continuum.
 % end of PACS codes
} %end of abstract
\maketitle
\section{Introduction}
\label{intro}
The use of QCD sum rules in studies investigating the properties of baryons has already a long 
history. Since the seminal papers of Shifman \textit{et al.} \cite{Shifman}, in which the 
QCD sum rule method was developed and the subsequent 
first application to baryonic channels by Ioffe \cite{Ioffe}, the sum rules of 
the nucleon have been continuously improved by including 
higher orders in the perturbative Wilson coefficients \cite{Krasnikov,Chung2,Jamin,Ovchinnikov,Shiomi,Sadovnikova} 
or non-perturbative 
power corrections \cite{Belyaev,Chung2,Leinweber1}. This even lead to attempts to determine the mass difference 
between the neutron and the proton \cite{Yang}, which certainly is a very difficult task because 
of the smallness of this difference compared to hadronic scales. 
Another important development was initiated in a paper by Leinweber \cite{Leinweber2}, 
in which, among other technical points, the choice of the interpolating field made by Ioffe 
was criticized and a new statistical method for the analysis of QCD sum rules was 
introduced. 
Furthermore, QCD sum rules also have been used to investigate the nucleon properties 
in nuclear matter \cite{Hatsuda,Cohen,Drukarev} or at finite temperature \cite{Adami}. 

However, in all these studies it was necessary to model the spectral function according to 
some specific functional form, the ``pole + continuum"-ansatz being the most popular one. 
Such a procedure inevitably incorporates strong assumptions on the spectral function 
into the analysis. 
This 
strategy works well when the actual spectral function has some resemblance with the 
chosen model, but will of course fail if this is not the case. For instance, as is 
known from studies using both QCD sum rules and lattice QCD, certain linear combinations 
of interpolating fields, which in principle carry the quantum numbers of the nucleon, 
couple to the nucleon ground state only very weakly. A QCD sum rule analysis of such 
interpolating fields, which uses the ``pole + continuum" ansatz, can only lead to 
ambiguous results. The problems become even more severe in studies of the spectral 
function at finite temperature or density, as the validity of the phenomenological 
"pole + continuum" ansatz for the spectral function in such an environment becomes less 
certain, rendering it more difficult to make educated 
guesses about its actual form. 

To avoid the above mentioned problems, and to reduce the assumptions that have to 
be made on the functional form of the spectral function, 
we have developed a novel method of 
analyzing QCD sum rules, which employs the maximum entropy method (MEM) \cite{Gubler1}. 
The method, which is based on Bayesian probability theory, was shown to give 
reasonable results in the analysis of the sum rule of the $\rho$-meson channel. 
Moreover, recently this approach was successfully applied to the charmonium sum rules at 
both zero and finite temperatures \cite{Gubler2}. 

In this paper, it is our main purpose to examine if and to what extent the MEM analysis 
can be applied to the sum rule of the nucleon. Throughout our investigations, we found 
that the MEM analysis of the Borel sum rule 
\begin{equation}
G_{\mathrm{OPE}}(M)  =\frac{1}{M^2}\int^{\infty} _{0} dt e^{-\frac{t}{M^2}}\rho (t), 
\label{eq:BorelSR}
\end{equation}
in fact fails to satisfactorily extract 
the nucleon spectral function in the ground state region. As we will discuss later in 
detail, this failure is mainly caused by the large contribution of the continuum to the 
OPE side of Eq.(\ref{eq:BorelSR}), which strongly deteriorates the contribution of the 
nucleon pole. 

On the other hand, the Gaussian sum rule \cite{Bertlmann,Orlandini}
\begin{equation}
G_{\mathrm{OPE}}(s,\tau)=\frac{1}{\sqrt{4\pi\tau}}\int^{\infty}_{0} dt e^{-\frac{(
t-s)^2}{4\tau}} \rho (t), 
\label{eq:GaussianSR}
\end{equation}
which, for the nucleon, is formulated for the first time in this paper, 
turns out to give better results and allows us to resolve the nucleon pole from 
the continuum. There are essentially two reasons for the superiority of the sum 
rule of Eq.(\ref{eq:GaussianSR}). First of all, the kernel of Eq.(\ref{eq:GaussianSR}), 
a Gaussian centered at $s$ with a width of $\sqrt{2\tau}$, 
collects more information on the spectral function $\rho(t)$ than the one in 
Eq.(\ref{eq:BorelSR}) when the integration over $t$ is carried out. This is 
especially true for small values of the width $\tau$, which we, however, 
cannot take arbitrarily small because below a certain threshold, the 
convergence of the operator product expansion (OPE) becomes poor. A similar 
situation occurs in the Borel sum rules, where the 
Borel mass is restricted from below due to the OPE convergence. 
The second reason also can be related to the kernel of Eq.(\ref{eq:GaussianSR}), 
containing two parameters $s$ and $\tau$, which can be freely chosen as long 
as the OPE converges. This freedom allows us to vary two parameters at the same time in the MEM
 analysis of \cite{Gubler1},  
leading to reasonable results for the extracted spectral function. 
Similar experiences also have been made in nuclear structure studies, where 
the Lorentz kernel has proven to be useful \cite{Efros,Efros2}. 

Furthermore, using the MEM analysis of the Gaussian sum rules, we are able to 
extract the spectral function not only in cases where the interpolating field 
strongly couples to the nucleon ground state and thus the ``pole + continuum" 
ansatz should be valid, but also in cases where only higher energy states contribute 
to the sum rules and hence the conventional analysis most likely fails to give meaningful 
results. This advantage is especially useful for examining which kind of interpolating 
field couples strongly to the nucleon ground state and is thus a suitable interpolating field 
for the  analysis, 
a question with a long and 
controversial history in QCD sum rules studies \cite{Ioffe,Chung1,Chung3,Ioffe2,Dosch,Leinweber2}. 
Our MEM analysis of the general 
operator given in Eq.(\ref{eq:interpolating field}) strongly suggests that 
the nucleon ground state only couples to $\eta_1(x)$ ($\beta=0$) and not to 
$\eta_2(x)$ ($\beta=\infty$)(see Eqs.(\ref{eq:eta1}) and (\ref{eq:eta2}) in Sec. \ref{II}). 
In addition, we have obtained some hints of excited states coupling to $\eta_2(x)$. 
These issues will be discussed in detail in Section \ref{subsection V3}.

The paper is organized as follows. In Sec. \ref{II}, the details of the Borel and Gaussian sum rule 
for the nucleon are explained. Next, the maximum entropy method (MEM) is elucidated in Sec. \ref{III}, 
after which in Sec. \ref{IV}, the results of the analysis using the Borel sum rule in combination with MEM are presented. 
In Sec. \ref{V}, the results of the analysis for the Gaussian sum rule are outlined, and 
in Sec. \ref{subsection V3}, the differences of the obtained spectral functions depending on the choice of the interpolating fields are discussed. 
Finally, the summary and conclusions are given in Sec. \ref{VI}.

\section{QCD sum rules for the nucleon}
\label{II}
The method of QCD sum rules is used to carry out 
the analysis of the spectral function as follows. First, 
we choose an interpolating field which has the quantum numbers of the nucleon, 
then its correlation function is calculated in the 
deep Euclidean 4-momentum region. 
Alternatively, the same correlation function at the physical 4-momentum region is expressed by 
the spectral function of the nucleon channel. 
The sum rules can then be constructed by 
equating the two expressions using a dispersion relation. 

For the nucleon, there are two independent local interpolating operators, 
\begin{equation}
\eta _{1}(x)=\epsilon ^{abc}(u^{Ta}(x)C\gamma_{5} d^{b}(x))u^{c}(x),
\label{eq:eta1}
\end{equation}
\begin{equation}
\eta _{2}(x)=\epsilon ^{abc}(u^{Ta}(x)Cd^{b}(x))\gamma_{5} u^{c}(x). 
\label{eq:eta2}
\end{equation}
Here, $abc$ are color indices, $C$ is the charge conjugation matrix and $T$ stands for the transposition operation. 
The spinor indices are omitted for simplicity. 
A general interpolating operator can thus be expressed as 
\begin{equation}
\eta(x)=\eta_{1}(x)+\beta\eta_{2}(x),
\label{eq:interpolating field}
\end{equation}
where $\beta$ is a real parameter. 
Here, the case of $\beta = -1$ is identified as the so-called ``Ioffe current" \cite{Ioffe}, which is 
often used in QCD sum rule studies of the nucleon. 

Using this interpolating operator, we define the correlation function as 
\begin{equation}
\begin{split}
\Pi(q) &= i\int e^{iqx}\langle 0|T[\eta (x)\overline{\eta}(0)]|0\rangle d^{4}x \\
       &=q\hspace{-.50em}/\Pi_{1}(q^2)+\Pi_2(q^2).
\label{eq:correlation function0}
\end{split}
\end{equation}
The imaginary part of $\Pi_1(q^2)$ satisfies the positivity condition, 
$\mathrm{Im}\Pi_1(q^2) \geq 0$, while $\mathrm{Im}\Pi_2(q^2)$ is not necessarily positive 
due to contributions of negative parity states. 
The positivity condition is, however, essential for the 
application of the MEM method and we thus consider only  
$\Pi_{1}(q^2)$ in the following. 
To make use of the information contained in $\mathrm{Im}\Pi_2(q^2)$, one would have 
to analyze the parity projected sum rules \cite{Jido}, which we plan to investigate in 
the future. 
In the deep Euclidean region ($-q^2 \to \infty$), $\Pi_{1}(q^2)$ can be calculated by using the 
operator product expansion (OPE). Including operators up to dimension 8, we get
\begin{equation}
\begin{split}
\Pi_{1}(q^2) &= -\frac{5+2\beta +5\beta ^2 }{128(2\pi ) ^4} q^{4} \ln (-q^2)\\
        &\quad -\frac{5+2\beta +5\beta ^2 }{256(2\pi ) ^2}
\langle \frac{\alpha _s}{\pi }G^2\rangle \ln (-q^2)  \\
        &\quad -\frac{7-2\beta -5\beta ^2 }{24} \langle \overline{q}q\rangle ^2\frac{1}{q^2} \\
        &\quad -\frac{13-2\beta -11\beta ^2 }{96} \langle \overline{q}q\rangle 
\langle \overline{q}g\sigma \cdot Gq\rangle 
\frac{1}{q^4}\\
        &\equiv \Pi_{\mathrm{OPE}}(q^2) 
\label{eq:ope}  
\end{split}
\end{equation}      
To obtain Eq.(\ref{eq:ope}), several approximations have been implemented. 
Firstly, only the lowest order in $\alpha _{s}$ is taken into account.
The validity of this approximation is not obvious, because
it is known that the first order $\alpha _{s}$ corrections are significant and 
lead to a considerable increase of the continuum contribution \cite{Leinweber2}. 
Nevertheless, our main goal of this paper 
is to examine whether the MEM analysis can be applied to the nucleon sum rule or not, and 
we thus ignore the $\alpha _{s}$ corrections here. 
For a more quantitative future analysis, the 
higher order corrections should certainly be taken into account. 
The second approximation arises from the use of the vacuum saturation, 
by which $\langle \overline{q}q \overline{q}q \rangle$ and 
$\langle \overline{q}q \overline{q}g\sigma \cdot Gq \rangle$ 
can be formally reduced to $\langle \overline{q}q \rangle^2$ and 
$\langle \overline{q}q\rangle \langle \overline{q}g\sigma \cdot Gq \rangle$, respectively. 
Although this approximation can be justified in the large 
$N_c$ limit \cite{Shifman}, it is not clear to what extent it is trustable at $N_c = 3$. 
Nonetheless, for the present qualitative analysis, 
we will assume this approximation to be valid. 

As already mentioned, $\Pi_{1}(q^2)$ can also be expressed in terms of the 
physical spectral function using the dispersion relation: 
\begin{equation}
\begin{split}
\Pi_{1}(q^2)&=\frac{1}{\pi}\int^\infty _{0} \frac{\mathrm{Im}\Pi _{1}(t)}{t-q^2}dt
=\int^\infty _{0} \frac{\rho (t)}{t-q^2}dt \\
&\equiv  \Pi _{\rho }(q^2),
\label{eq:disp}
\end{split}
\end{equation}
where the definition Im$\Pi_1(t)=\pi\rho(t)$ is used for the spectral function. 
Our goal is now to extract $\rho(t)$ from the sum rule obtained 
by equating Eq.(\ref{eq:ope}) and Eq.(\ref{eq:disp}).
It should be noted here that subtractions are necessary in order to 
make the integral of Eq.(\ref{eq:disp}) convergent. 
In the case of the nucleon, the subtraction terms are 
$\Pi _{1}(0) + \Pi _{1}'(0)q^{2} + \frac{\Pi _{1}''(0)}{2} q^{4}$, 
which will disappear after 
transforming Eq.(\ref{eq:disp}) into the Borel or Gaussian sum rules. 
How this is done will be explained in the following subsections. 

\subsection{Borel sum rule}
\label{subsection Borel sum rule}
In the case of the Borel sum rule, we transform $\Pi_{1}(q^2)$ using the Borel transformation
$\hat B_{M }$, defined below: 
\begin{equation}
\hat B_{M }= \lim_{\genfrac{}{}{0pt}{}{-q^{2},n \to \infty}{-q^2/n=M^2}}
\frac{(-q^2)^n}{(n-1)!} (\frac{d}{dq^2})^n.
\label{eq:Borel transformation_M}
\end{equation}
Applying $\hat B_{M }$ to Eq.(\ref{eq:ope}), 
we get the following expression 
for $G_{\mathrm{OPE}}(M)\equiv \hat B_{M}\Pi_{\mathrm{OPE} }(q^2)$: 
\begin{equation}
\begin{split}
G_{\mathrm{OPE}}(M) &= \frac{5+2\beta +5\beta ^2 }{64(2\pi ) ^4} M^4 \\
        &\quad +\frac{5+2\beta +5\beta ^2 }{256(2\pi ) ^2}\langle \frac{\alpha _s}{\pi }G^{2}\rangle   \\
        &\quad +\frac{7-2\beta -5\beta ^2 }{24} \langle \overline{q}q\rangle ^2\frac{1}{M^2} \\
        &\quad -\frac{13-2\beta -11\beta ^2 }{96} 
\langle \overline{q}q\rangle \langle \overline{q}g\sigma \cdot Gq\rangle 
\frac{1}{M^4} .
\label{eq:opeBorel}
\end{split}
\end{equation}
Meanwhile, applying the Borel transformation to Eq.(\ref{eq:disp}), 
we obtain for $G_{\rho }(M)\equiv \hat B_{M}\Pi_{\rho }(q^2)$: 
\begin{equation}
G_{\rho}(M)  =\frac{2}{M^2}\int^\infty _0 e^{-\frac{\omega ^2}{M^2}}\omega\rho (\omega)d\omega .
\end{equation}
Here , $t=\omega^2$ was used. 
Note that at this point, the subtraction terms have been eliminated and 
the integral of the right hand side converges. 
This leads us to the final form of the Borel sum rule, 
\begin{equation}
G_{\mathrm{OPE}}(M)=G_{\rho} (M) =\frac{2}{M^2}\int^\infty _0 
e^{-\frac{\omega ^2}{M^2}}\omega\rho (\omega)
d\omega .
\label{eq:Borel SR}
\end{equation}

\subsection{Gaussian sum rule}
\label{subsection Gaussian sum rule}
The Gaussian sum rule, first introduced in \cite{Bertlmann},
exhibits another way of improving Eq.(\ref{eq:disp}).
Based on the idea of local duality, it provides a formulation
for the convolution of the spectral function with a Gaussian
kernel. As this sort of sum rule is not often discussed in the
literature and the specific case of the nucleon has to
our knowledge not even been formulated, we will explain
each step in some detail, following closely the formulation
given in \cite{Orlandini}. 
 
Going to the complex plane of $q^2$ and taking the difference between 
$\Pi_{1}(q^{2}=s+i\Delta)$ and $\Pi_{1}(s-i\Delta)$ of Eq.(\ref{eq:disp}), 
where $s$ and $\Delta$ are real, we obtain  
\begin{equation}
\frac{\Pi _{1}(s+i\Delta)-\Pi _{1}(s-i\Delta)}{2i \Delta}=
\int^\infty _{0} \frac{\rho (t)}{(t-s)^2 + \Delta^2}dt.
\end{equation}
At this stage, the integral above is not convergent and the subtraction terms are not yet 
fully eliminated. 
Applying the following Borel transform
$\hat B_{4\tau }$: 
\begin{equation}
\hat B_{4\tau }= \lim_{\genfrac{}{}{0pt}{}{\Delta^{2},n \to \infty}{\Delta^2/n=4\tau}}
\frac{(-\Delta^2)^n}{(n-1)!} (\frac{d}{d\Delta^2})^n ,
\end{equation}
and using 
\begin{equation}
\hat B_{4\tau }[\frac{1}{(t-s)^2 + \Delta^2}]
=\frac{1}{4\tau}e^{-\frac{(t-s)^2}{4\tau}},
\label{eq:Borel formula1}
\end{equation}
we get
\begin{equation}
\begin{split}
2\sqrt{\frac{\tau}{\pi}}
\hat B_{4\tau }(&\frac{\Pi _{1}(s + i\Delta)-\Pi _{1}(s - i\Delta)}{2i\Delta})\\
&=
\frac{1}{\sqrt{4\pi\tau}}\int^{\infty}_{0} dt e^{-\frac{(t-s)^2}{4\tau}} \rho (t ) \\
&\equiv    G_{\rho}(s,\tau).
\label{eq:Gaussian dummy1}
\end{split}
\end{equation}
Here, the subtraction terms have disappeared and the integral in the above equation 
is convergent. 

As a next step, we will now show that Eq.(\ref{eq:Gaussian dummy1}) can also be rewritten by using 
the inverse Laplace transform $\hat L^{-1}$. 
\begin{figure*}[!htbp]
 \begin{center}
  \includegraphics[scale=1.3]{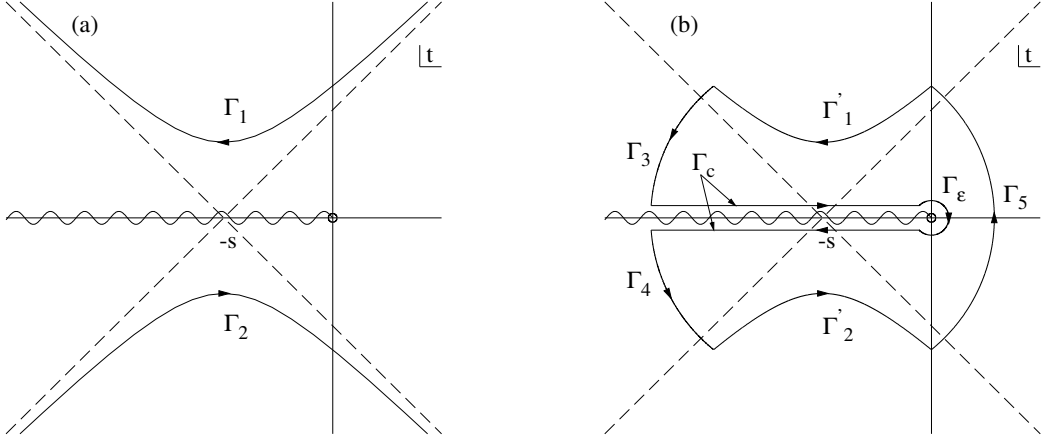}
 \end{center}
 \caption{
(a) Contour of integration $\Gamma _{1} + \Gamma _{2}$ in 
Eq.(\ref{eq:Gaussian dummy2}). 
The branch cut of $\Pi _{1}(-t)$ is represented  by the wavy line on the negative real axis. 
(b) Closed contour $C(R)$ of Eq.(\ref{eq:Gaussian dummy4}). 
The outer lines 
$\Gamma _{3}$, $\Gamma _{4}$ and $\Gamma _{5}$ are a part of the circle of radius $R$ centered at $-s$, 
while $\Gamma_{\epsilon}$ denotes a circular line with 
radius $\epsilon$, centered at the origin. 
The branch cut of $\Pi _{1}(-t)$ is represented  by the wavy line on the negative real axis and 
the contour $\Gamma _{c}$ is 
the straight line of the contour above and below the 
branch cut. The contours $\Gamma ^{'}_{1}$ and $\Gamma ^{'}_{2}$ are parts of $\Gamma_{1}$ and  
$\Gamma_{2}$, respectively.}
 \label{fig:contour1}
\end{figure*}
Using 
\begin{equation}
\begin{split}
\frac{1}{(t-s)^2 + \Delta^2}&= \hat L[e^{-x(t-s)^2}]\nonumber\\ 
&= \int^\infty _{0}e^{-\Delta^2x}e^{-x(t-s)^2}dx,
\end{split}
\end{equation} 
the kernel $e^{-x(t-s)^2}$ can be re-expressed by $\hat L^{-1}$: 
\begin{equation}
\begin{split}
e^{-x(t-s)^2}&=\hat L^{-1}[\frac{1}{(t-s)^2 + \Delta^2}] \\
&= \frac{1}{2\pi i}\int^{b+i\infty} _{b-i\infty}e^{\Delta^2 x}
\frac{1}{(t-s)^2 + \Delta^2}d\Delta^2.
\label{eq:Laplace formula1}
\end{split}
\end{equation}
Setting $x = \frac{1}{4\tau}$, 
the left-hand side of Eq.(\ref{eq:Gaussian dummy1}) can thus be rewritten as 
%\begin{widetext}
\begin{equation}
\begin{split}
&\frac{1}{\sqrt{4\pi \tau}}\hat L^{-1}
[\frac{\Pi _{1}(s + i\Delta)-\Pi _{1}(s - i\Delta)}{2i \Delta}] \\
&= \frac{1}{\sqrt{4\pi \tau}} \frac{1}{2\pi i}\int^{b+i\infty} _{b-i\infty}
[\frac{\Pi _{1}(s + i\Delta)-\Pi _{1}(s - i\Delta)}{2i\Delta}]
e^{\frac{\, \Delta^{2}}{4\tau }}d\Delta^2.\nonumber
\end{split}
\end{equation}
%\end{widetext}
Then, replacing $s + i\Delta$ by $-t$ in the first and $s - i\Delta$ by $-t$ 
in the second term, we get
\begin{equation}
G_{\rho }(s,\tau )=\frac{1}{\sqrt{4\pi \tau}}\frac{1}{2\pi i}\int _{\Gamma _{1}+\Gamma _{2} }
\Pi _{1}(-t)\mathrm{exp} (-\frac{(t+s)^2}{4\tau})dt
\label{eq:Gaussian dummy2}
\end{equation}
where the contour $\Gamma _{1}+\Gamma _{2}$ is shown in Fig. \ref{fig:contour1}a. 

Next, to obtain a sum rule that is practically usable, 
we consider the contour $C(R)$ shown in Fig. \ref{fig:contour1}b. 
Taking the limit $R \rightarrow \infty$ and $\varepsilon \rightarrow 0$, 
we are lead to the equation given below: 
\begin{equation}
\begin{split}
&G_{\rho }(s,\tau )\\
&=-\lim_{R\rightarrow \infty }\frac{1}{\sqrt{4\pi \tau}}
\frac{1}{2\pi i}\int _{\Gamma _{c}+\Gamma _{\epsilon } }
\Pi _{1}(-t)\mathrm{exp} (-\frac{(t+s)^2}{4\tau})dt. 
\label{eq:Gaussian dummy4}
\end{split}
\end{equation}
Substituting the right-hand side of Eq.(\ref{eq:ope}) into Eq.(\ref{eq:Gaussian dummy4}) 
and examining the various terms, we see that 
the perturbative and dimension four terms 
only give contributions on the contour $\Gamma_{c}$. 
Meanwhile, the dimension six and eight terms 
do not have a branch discontinuity, 
but a pole at $t=0$ and therefore only contribute on $\Gamma_{\epsilon}$. 
Using
\begin{equation}
\int^{0}_{-\infty }\mathrm{exp} (-\frac{(t+s)^2}{4\tau })dt
=\sqrt{\tau \pi}[1+\mathrm{erf} (\frac{s}{2\sqrt{\tau}})]
\end{equation}
where the error function $\mathrm{erf}(x)$ is defined as 
\begin{equation}
\mathrm{erf}(x)=\frac{2}{\sqrt{\pi}}\int^x_0 e^{-x^2}dx
\end{equation}
and 
\begin{equation}
\begin{split}
-\frac{1}{2\pi i}\int_{\Gamma _{\epsilon}} \frac{1}{t^n}\mathrm{exp}(-\frac{(t+s)^2}{4\tau })dt \\
=\lim_{t \rightarrow 0}\frac{1}{(n-1)!}\frac{d^{n-1}}{dt^{n-1}}\mathrm{exp}(-\frac{(t +s)^{2}}{4\tau})
\\ \quad (n=1,2, \dots), 
\end{split}
\end{equation}
we obtain 
\begin{equation}
\begin{split}
G_{\mathrm{OPE}}&(s, \tau ) = \frac{5+2\beta +5\beta ^2 }{128(2\pi ) ^4}\\ 
& \times [\tau (1+\frac{s^2}{2\tau})[1+\mathrm{erf}(\frac{s}{2\sqrt{\tau}})] 
+ \frac{s\sqrt{\tau}}{\sqrt{\pi}} \mathrm{exp}(-\frac{s^2}{4\tau})] \\
 & + \frac{5+2\beta +5\beta ^2 }{512(2\pi ) ^2}\langle \frac{\alpha _s}{\pi }G^2\rangle 
[1+\mathrm{erf}(\frac{s}{2\sqrt{\tau}})]  \\
 & +\frac{1}{\sqrt{\pi\tau}}\frac{7-2\beta -5\beta ^2 }{48} \langle \overline{q}q\rangle ^{2}
\mathrm{exp}(-\frac{s^2}{4\tau}) \\
 & +\frac{1}{\sqrt{\pi\tau}}\frac{13-2\beta -11\beta ^2 }{384} 
\langle \overline{q}q\rangle \langle \overline{q}g\sigma \cdot Gq\rangle \frac{s}{\tau}\mathrm{exp}(-\frac{s^2}{4\tau}). 
\label{eq:opegaussian}
\end{split}
\end{equation}
This then finally leads to the following form of the Gaussian sum rule, 
from which information of the spectral function $\rho(\omega)$ can be extracted: 
\begin{equation}
G_{\mathrm{OPE}}(s,\tau) =G_{\rho}(s,\tau)
= \frac{2}{\sqrt{4\pi\tau}}\int^{\infty}_{0} d\omega \cdot \omega 
e^{-\frac{(\omega^2-s)^2}{4\tau}} \rho (\omega ). 
\label{Gaussian SR}
\end{equation}
Here, we have again set $t=\omega^2$.

\section{Maximum Entropy Method}
\label{III}
In this section, we will briefly explain the maximum entropy method (MEM). 
The general sum rule that we aim to analyze with this method is 
\begin{equation}
G_{\mathrm{OPE}}(x)=G_{\rho}(x)=\int^\infty _0 d\omega K(x,\omega )\rho (\omega), 
\label{eq:sr} 
\end{equation}
where $K(x,\omega)$ stands for the kernel of either the Borel or the Gaussian 
sum rule, and $x$ denotes the Borel mass $M$ in the Borel sum rule or $(s,\tau)$ in the 
Gaussian sum rule, respectively. 
Using Eq.(\ref{eq:sr}), we extract $\rho (\omega)$ 
from $G_{\mathrm{OPE}}(x)$. 
However, this is nontrivial, because $G_{\mathrm{OPE}}(x)$ contains a certain error $\sigma(x)$ 
arising from the uncertainty of the condensate parameters and 
furthermore can only be used in the region where 
the OPE converges. For a specific value of $x$, $G_{\rho}(x)$ is only 
sensitive to a limited region of $\rho(\omega)$, making it hard to obtain 
$\rho(\omega)$ in a wide range of $\omega$. 
Because of all these problems, rigorously solving Eq.(\ref{eq:sr}) is in fact 
an ill-posed problem. Neverthless, the MEM technique enables us at least to statistically 
determine the most probable form of $\rho(\omega)$. 
Using MEM has moreover the advantage that the solution of $\rho(\omega)$ is unique if it exists \cite{Asakawa}. 
Therefore, different from the usual $\chi ^{2}$ fitting, the problem of local minima does not occur. 

Let us now discuss the connection between MEM and Bayes' theorem. 
First, 
we define H to denote the prior knowledge on 
$\rho (\omega )$ such as positivity and asymptotic values. 
Then, $P[\rho |GH]$, the conditional probability of $\rho$ 
given $G_{\mathrm{OPE}}$ and H is rewritten using Bayes' theorem as 
\begin{equation}
P[\rho |GH]=\frac{P[G|\rho H]P[\rho |H]}{P[G|H]}.
\label{eq:MEMdummy1}
\end{equation}
Here, $P[\rho |H]$ is the so-called prior probability, and $P[G|\rho H]$ stands for the 
likelihood function.
The most probable form of $\rho(\omega)$ is obtained, by maximizing $P[\rho |GH]$ of Eq.(\ref
{eq:MEMdummy1}). 
For carrying out this task, one needs the concrete functional forms of 
$P[\rho |H]$ and $P[G|\rho H]$, which can be expressed as \cite{Gubler1,Asakawa}   
\begin{equation}
\begin{split}
P[G|\rho H]&=e^{-L[\rho ]},\\
P[\rho |H]&=e^{\alpha S[\rho ]}, 
\end{split}
\end{equation}
where $\alpha$ is a real positive number and 
the functionals $L[\rho]$ and $S[\rho]$ are defined as 
\begin{equation}
\begin{split}
L[\rho] =\frac{1}{2(x_{max}-x_{min})}\int^{x_{max}}_{x_{min}}dx
\frac{[G_{ope}(x)-G_{\rho }(x)]^2}{\sigma ^{2}(x)},
\label{eq:likelihood cont}
\end{split}
\end{equation}
and 
\begin{equation}
S[\rho] =\int^{\infty}_{0} d\omega [\rho  (\omega ) -m(\omega ) - \rho (\omega ) 
\log(\frac{\rho (\omega )}{m(\omega )})],
\label{eq:entropy cont}
\end{equation}
where, the error $\sigma(x)$ in Eq.(\ref{eq:likelihood cont}) is evaluated by the statistical method explained 
in \cite{Leinweber2}. 
$S[\rho]$ is also known as the Shannon-Jaynes entropy and the positive 
function $m(\omega)$ appearing in this expression is called the default model. 

Since $P[G|H]$ does not depend on $\rho (\omega )$, we drop it 
as it corresponds only to a normalization constant. 
Hence, $P[\rho |GH]$ is given as 
\begin{equation}
\begin{split}
P[\rho |GH]&\propto P[G|\rho  H]P[\rho |H]\nonumber\\
 &= e^{Q[\rho ]},
\end{split}
\end{equation}
where, 
\begin{equation}
Q[\rho ]= \alpha S[\rho ]-L[\rho ],              
\end{equation}
which means that to get the most probable $\rho(\omega)$, one has to solve the numerical problem of obtaining the form of 
$\rho(\omega)$ that maximizes Q[$\rho $], 
for which we use the Bryan algorithm \cite{Bryan}. 
Once a $\rho _{\alpha } (\omega )$ that maximizes Q$[\rho]$ for a fixed value of $\alpha$ is found, 
this parameter is integrated out by averaging $\rho $ over a range of values of $\alpha$ and 
assuming that $P[\rho |GH]$ is sharply peaked around its maximum $P[\rho _{\alpha} |GH]$: 
\begin{equation}
\begin{split}
\rho _{\mathrm{out}} (\omega ) = \int [d\rho ] \int d\alpha \rho (\omega )
P[\rho |GH] P[\alpha |GH] \\
\simeq \int d\alpha \rho _{\alpha}(\omega ) P[\alpha |GH]
\label{eq:alphaaverage}
\end{split}
\end{equation}
The integrated range of $\alpha$ is $P[\rho _{\alpha} |GH] \geq 0.1\times 
P[\rho _{\alpha _{\mathrm{max}}} |GH]$ 
where $\alpha _{\mathrm{max}}$ is the value of $\alpha $ which maximizes $P[\rho _{\alpha} |GH]$. 
By using Bayes' theorem, $P[\alpha |GH]$ is expressed as (for details see 
\cite{{Asakawa}}): 
\begin{equation}
P[\alpha |GH] \propto P[\alpha |H] \exp ( \frac{1}{2} \sum_k \log \frac{\alpha}{\alpha + \lambda_k} 
+ Q[\rho _{\alpha }]), 
\nonumber
\end{equation}
where $\lambda_k$ are the eigenvalues of the matrix, 
\begin{equation}
\Lambda_{ij} = \sqrt{\rho _{i}} \frac{\partial ^2 L}{\partial \rho _{i} \partial \rho _{j}} 
\sqrt{\rho _{j}} \bigg|_{\rho =\rho _{\alpha }}, 
\nonumber
\end{equation}
and $\rho _{i}$ denotes $\rho (\omega _{i})\, d\omega$. 

In the numerical analysis, Eqs.(\ref{eq:sr},\ref{eq:likelihood cont},\ref{eq:entropy cont}) 
are discretized as follows: 
\begin{equation}
G(x_{i} ) = \sum _{\omega _{j} = \omega _{\mathrm{min}}} ^{\omega _{\mathrm{max}}}d\omega 
K(x_{i}, \omega _{j}) \rho (\omega _{j} ), 
\label{eq:Gaussian SR dis}
\end{equation}
\begin{equation}
L(\rho ) = \frac{1}{2(x_{\mathrm{max}}-x_{\mathrm{min}})}
\sum ^{x_{\mathrm{max}}}_{x_{i} = x_{\mathrm{min}}}dx
\frac{[G_{\mathrm{OPE}}(x_{i})-G_{\rho }(x_{i})]^2}{\sigma ^{2}(x_{i})},
\label{eq:lilelihood dis}
\end{equation}
\begin{equation}
\begin{split}
S(\rho )&= \sum _{\omega _{j} =\omega _{\mathrm{min}}} ^{\omega _{\mathrm{max}}}d\omega 
[\rho (\omega _{j} ) -m(\omega _{j} ) - \rho (\omega _{j} ) 
\log(\frac{\rho (\omega _{j} )}{m(\omega _{j})})].
\label{eq:entropy dis}
\end{split}
\end{equation}
In Eqs.(\ref{eq:Gaussian SR dis} - \ref{eq:entropy dis}),
we take 100 data points in the analyzed region of the variable $x$
($N_{x}$=100) and 600 data points for the variable $\omega$ 
($N_{\omega}$=600). Furthermore, we adjust 
$\omega _{\mathrm{min}}$, $\omega _{\mathrm{max}}$, $d\omega$ and $dx$ to 0[GeV], 6[GeV], 
$(\omega_{\mathrm{max}} - \omega_{\mathrm{min}})/N_ {\omega}$ and 
$(x_{\mathrm{max}} - x_{\mathrm{min}})/N_ {x}$, respectively. 

Finally, the method of evaluating the error of the obtained spectral function 
is introduced below. 
We estimate the averaged error of $\rho_{\mathrm{out} } (\omega )$ over a 
certain region $[\omega _{1} ,\omega _{2}]$.
$ \langle (\delta \rho _{\alpha  })^{2}\rangle _{\omega _{1} ,\omega _{2} }$, the dispersion of 
$\rho _{\alpha } (\omega )$ in $[\omega _{1} ,\omega _{2}]$ is defined as 
\begin{equation}
\begin{split}
& \langle (\delta \rho _{\alpha  })^{2}\rangle _{\omega _{1} ,\omega _{2} } \\
&= \frac{1}{(\omega _{2} -\omega _{1} )^{2}}
\int [d\rho ] \int _{\omega _{1} }^{\omega _{2} }d\omega d\omega ' 
\delta \rho _{\alpha  } (\omega) \delta \rho _{\alpha  } (\omega ' ) 
\mathrm{P}[\rho |\mathrm{GH}] \nonumber\\
&= - \frac{1}{(\omega _{2} -\omega _{1} )^{2}} \int _{\omega _{1} }^{\omega _{2} }d\omega d\omega ' 
(\frac{\delta ^{2}Q}{\delta \rho (\omega ) \delta \rho (\omega ' )})^{-1} |_{\rho = \rho _{\alpha}}, 
\end{split}
\end{equation}
where $\delta \rho _{\alpha  } (\omega ) = \rho (\omega ) - \rho _{\alpha  } (\omega )$ and 
the gaussian approximation  of $\mathrm{P}[\rho |\mathrm{GH}]$ around $\rho_{\alpha}$ has been used. 
After averaging over $\alpha$ in the same way as in Eq.(\ref{eq:alphaaverage}), we arrive at the 
final value of $\delta \rho _{\mathrm{out}}$: 
\begin{equation}
 \langle \delta \rho _{\mathrm{out}}\rangle _{\omega_1,\omega_2}
 = \int _{\alpha _{\mathrm{min}}}^{\alpha _{\mathrm{max}}} d\alpha  
\sqrt{\langle (\delta \rho _{\alpha  })^{2}\rangle _{\omega_1,\omega_2}} \mathrm{P}[\alpha |\mathrm{GH}]. 
\end{equation}
The errors of the spectral functions shown in the figures of this paper 
were all estimated by this method. 
These errors will be indicated by three horizontal lines, whose lengths 
stand for the region $[\omega _{1} ,\omega _{2}]$, over which the average is 
taken, while their heights correspond to 
$\langle \rho_{\mathrm{out}} \rangle + \langle \delta \rho_{\mathrm{out}} \rangle$, 
$\langle \rho_{\mathrm{out}} \rangle$, 
$\langle \rho_{\mathrm{out}} \rangle - \langle \delta \rho_{\mathrm{out}} \rangle$, 
respectively. 
For more details about MEM, we refer the reader to \cite{Asakawa,Jarrel}. For an
application of this method to the nucleon channel in the framework of
lattice QCD, see \cite{Sasaki} and for a discussion of specific issues
related to QCD sum rules, consult \cite{Gubler1}.

\section{Analysis using the Borel sum rule}
\label{IV}
In this section, we will analyze the nucleon spectral function for the Borel sum rule. 
It is easily understood from dimensional considerations that unlike in the meson case, the contribution of the continuum states to the 
baryon spectral function is proportional to $\omega^4$ and thus 
strongly enhanced. As was done in similar studies using MEM 
and lattice QCD, we will therefore analyze $\rho(\omega)$/$\omega^4$ instead of $\rho(\omega)$ and 
hence from now on denote $\rho(\omega)$/$\omega^4$ as $\rho(\omega )$, 
leading to the equations below: 
\begin{equation}
\begin{split}
G_{\mathrm{OPE}}(M)= G_{\rho }(M)&=\int^\infty _0d\omega K(M,\omega )\rho(\omega), \\
K(M,\omega)&=\frac{2\omega^{5}}{M^2} e^{-\omega^{2}/M^2}. \nonumber
\end{split}
\end{equation}

\subsection{Analysis using mock data}
\label{subsection Analysis using mock data}
\begin{figure}[!tbp]
 \begin{center}
  \includegraphics[scale=0.6]{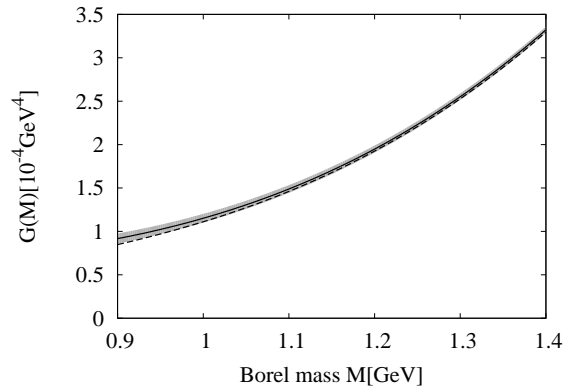}
 \end{center}
 \caption{Comparison of the mock data and the OPE data at $\beta=-1$. 
The solid line shows $G_{\mathrm{OPE}}(M)$ of Eq.(\ref{eq:opeBorel}), 
with the respective error indicated by the gray region. 
The dashed line denotes $G_{\mathrm{mock}}(M)$ for which the mock spectral function 
of Eq.(\ref{eq:mockspec}) was used.}
 \label{fig:comparing G(OPE)with G(rho)}
\end{figure}
In order to check the effectiveness of MEM to extract the spectral function of the nucleon, 
we first carry out an analysis using mock data. 
The employed mock spectral function is given below: 
\begin{equation}
\begin{split}
\rho _{\mathrm{mock}}(\omega )=\frac{\lambda ^{2}}{2M_{N} ^{5}}\delta (\omega - & M_{N}) +  
\frac{5+2\beta +5\beta ^2 }{128(2\pi ) ^4}
\frac{1}{1+e^{\frac{(\omega_0 -\omega )}{\delta}}},
\label{eq:mockspec}
\end{split}
\end{equation}
\begin{table}[t]
\begin{center}
\begin{tabular}{|c|c|c|c|}
\hline
$\langle \overline{q}q\rangle $&$\langle \frac{\alpha _s}{\pi }G^{2}\rangle $&
$\langle \overline{q}g\sigma \cdot Gq\rangle $/$\langle \overline{q}q\rangle $
\\ \hline
$-(0.24\pm0.01)^{3} \,\mathrm{GeV}^{3}$&$0.012\pm 0.0036 \,\mathrm{GeV}^{4} $&$0.8\pm 0.2 \,\mathrm{GeV}^{2} $
\\ \hline
\end{tabular}
\caption{Values of the parameters appearing in the OPE, 
taken from \cite{Colangelo}.}
\label{the condensate parameters}
\end{center}
\end{table}
\begin{figure*}[!tbp]
 \begin{center}
  \includegraphics[scale=0.9]{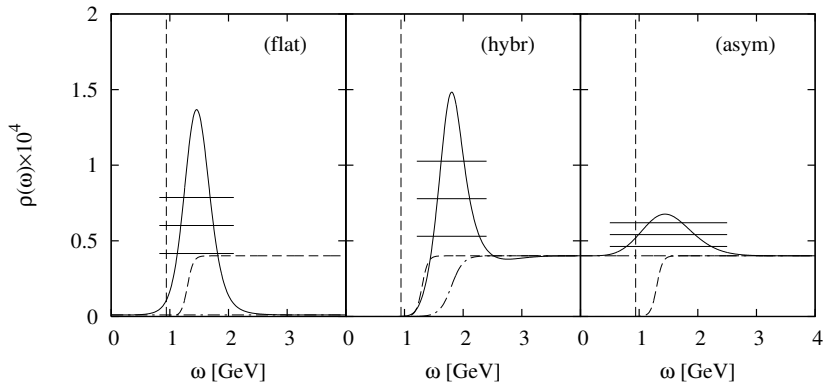}
 \end{center}
 \caption{Spectral functions extracted from $G_{\mathrm{mock}}(M)$ using various default models at $\beta=-1$. 
The analyzed Borel mass region is 0.91$\,$GeV $\leq $ $M$ $\leq $ 0.97$\,$GeV. 
The solid line shows $\rho(\omega)$, the dashed line depicts the input mock spectral function and the 
dash-dotted line stands for the default model. 
As we explain in Section 3, the three horizontal lines correspond to 
$\langle \rho_{\mathrm{out}} \rangle + \langle \delta \rho_{\mathrm{out}} \rangle$, 
$\langle \rho_{\mathrm{out}} \rangle$, 
$\langle \rho_{\mathrm{out}} \rangle - \langle \delta \rho_{\mathrm{out}} \rangle$, 
respectively. 
Similar horizontal lines in following figures are understood to have the same meaning. 
}
\label{fig:BorelSR_0.91_0.97_mock}
\end{figure*}
where we use the following values for the various parameters:  
\begin{equation}
\begin{split}
M_{N}&=940 \,\mathrm{MeV},\,\omega_{0}=1.3 \,\mathrm{GeV},\, \\ 
\delta&=0.05 \,\mathrm{GeV},\,\lambda ^{2}
=\frac{0.19}{(2\pi)^4} \,\mathrm{GeV}^6. 
\end{split}
\end{equation}
Here, $\rho _{\mathrm{mock}}(\omega )$ is constructed to have a narrow ground state pole 
and a continuum, which approaches the perturbative value at high energy. 
Defining now 
\begin{equation}
G_{\mathrm{mock}}(M) \equiv \int^\infty _0d\omega K(M,\omega )\rho _{\mathrm{mock}} (\omega ),
\end{equation}
we apply the MEM procedure to 
\begin{equation}
G_{\mathrm{mock}}(M) =G_{\rho }(M)=\int^\infty _0d\omega K(M,\omega)\rho (\omega ). 
\label{eq:integralmock}
\end{equation}
The residue of the nucleon pole $\lambda$ was fitted so that $G_{\mathrm{mock}}(M)$ matches 
$G_{\mathrm{OPE}}(M)$ at $\beta = -1$ in the analyzed Borel mass region. 
For comparison, 
$G_{\mathrm{OPE}}(M)$ and $G_{\rho}(M)$ are shown in Fig. \ref{fig:comparing G(OPE)with G(rho)}. 
The values of the parameters appearing in Eq.(\ref{eq:ope}) 
are shown in Table \ref{the condensate parameters}. 
As can be observed in Fig. \ref{fig:comparing G(OPE)with G(rho)}, 
$G_{\mathrm{OPE}}(M)$ and $G_{\rho}(M)$ are consistent within the range of the error, 
which shows that the ``pole + continuum" ansatz describes the OPE data well. 
However, this does not necessarily imply that the integral of Eq.(\ref{eq:integralmock}) 
can be reliably inverted and that valid information on the nucleon pole can be 
extracted. To investigate to what extent this is 
possible, we now analyze $G_{\mathrm{mock}}(M)$. 
For a realistic analysis, we here employ the error obtained from $G_{\mathrm{OPE}}(M)$, 
$\sigma_{\mathrm{mock}}(M) = \sigma_{\mathrm{OPE}}(M)$. 

To use MEM, we must at first fix the default model $m(\omega)$. 
A reasonable choice for $m(\omega)$ should reflect our prior knowledge on the spectral function 
such as the asymptotic behavior at low or high energy. 
To test several possible choices, we here introduce three types of default models.
The first one is a constant consistent only 
with the asymptotic behavior of the spectral function at low energy, therefore 
lying close to 0:
\begin{equation}
m_{\mathrm{flat}}(\omega)=1.0\times 10^{-6}.
\label{eq:default flat}
\end{equation}
The detailed value of $m_{\mathrm{flat}}$ is not so important, as long as 
it can be considered to be small enough compared to the 
asymptotic value at high energy. 
As we will discuss in more detail in the section dealing with the gaussian sum rule, 
we indeed have found that the position of the lowest lying pole and its residue of the 
obtained spectral functions 
depend on the value of $m_{\mathrm{flat}}$ only weakly. 
The second default model is also a constant 
which now reflects the asymptotic behavior at high energy:
\begin{equation}
m_{\mathrm{asym}}(\omega )=\frac{5+2\beta +5\beta ^2 }{128(2\pi ) ^4}.
\label{eq:default asym}
\end{equation}
The third one is a combination of the first two with the correct behavior at both 
high and low energy: 
\begin{equation}
\begin{split}
m_{\mathrm{hybr}}(\omega )&=\frac{5+2\beta +5\beta ^2 }{128(2\pi ) ^4} 
\frac{1}{1+e^{(\omega_0 -\omega )/\delta}}, \\
\omega_{0}&=1.8\,\mathrm{GeV}, \delta =0.1\,\mathrm{GeV}.
\label{eq:default hybr}
\end{split}
\end{equation}

In the following, the analysis is carried out at $\beta = -1$. 
In the case of the conventional method, which assumes the spectral function 
to have a specific functional form such as the 
``pole + continuum" ansatz, 
the analyzed Borel mass region is restricted 
so that the ratio of the highest dimensional term 
is less than 0.1 of the whole $G_{\mathrm{OPE}}(M)$ 
to have some confidence on the OPE convergence. 
The Borel mass region is further limited by the condition that 
the contribution from the continuum states 
in $G_{\rho }(M)$ should be less than 0.5 to make sure that the lowest pole 
dominates the sum rule. 
The Borel mass region determined according to these restrictions 
becomes 0.91$\,$GeV $\leq $ $M$ $\leq $ 0.97$\,$GeV. 
As pointed out in \cite{Leinweber2}, this region is very narrow and thus 
we expect that it will be very difficult to extract $\rho(\omega)$ in 
a wide range of $\omega$ with this small amount of available information. 
Although, when using MEM, 
we can in principle employ values of $M$ above 0.97$\,$GeV, 
we here first analyze the spectral function using the Borel mass region 0.91$\,$GeV $\leq $ $M$ $\leq $ 0.97$\,$GeV. 
The results are shown in Fig. \ref{fig:BorelSR_0.91_0.97_mock}. 

It is clear from Fig. \ref{fig:BorelSR_0.91_0.97_mock}, that the obtained lowest peaks 
lie much above the input value of 940$\,$MeV. Hence, as expected, we cannot extract much 
information on the nucleon pole from this sum rule. 
Especially, in the case of the default model $m_{\mathrm{asym}}(\omega )$, 
the spectral function in the low energy region approaches $m_{\mathrm{asym}}(0)$ 
and we can only observe a small lowest peak. 
A similar tendency was observed in the $\rho$-meson channel \cite{Gubler1}. 
We will therefore abandon this default model in the following. 
In the case of $m_{\mathrm{flat}}(\omega )$, 
although the high energy behavior wrongly approaches 0 at high energy, 
the low energy behavior, which is the main focus of our interest, seems to be reasonable. 
On the other hand, using $m_{\mathrm{hybr}}(\omega)$ leads to the correct behavior at 
both high and low energy. From these results, we can infer that the default model 
completely determines the asymptotic values of the spectral function. This behavior at 
high and low energy should therefore be considered to be an input in the current 
analysis. This can be understood from the properties of the kernel 
$K(M,\omega)$, leading to a function $G_{\rho}(M)$, which is insensitive to the 
values of $\rho(\omega)$ at large and small values of $\omega$. From the behavior of 
$K(M,\omega)$, we can also expect that increasing the upper boundary of $M$ allows the analysis 
to become sensitive to $\rho(\omega)$ at higher energy regions. 

\begin{figure}[tbp]
 \begin{center}
  \includegraphics[scale=0.68]{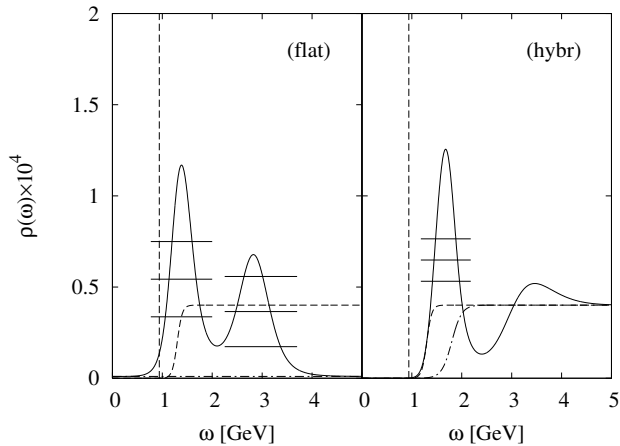}
 \end{center}
 \caption{Spectral functions extracted from $G_{\mathrm{mock}}(M)$ using various default models at $\beta=-1$. 
The analyzed Borel mass region is 0.91 GeV $\leq M \leq $ 1.4 GeV. 
The solid line shows $\rho(\omega)$, the dashed line depicts the input mock spectral function and the 
dash-dotted line stands for the default model.}
 \label{fig:BorelSR_0.91_1.4_mock}
\end{figure}

For investigating this case, 
we analyze the spectral function under the 
condition of $M_{\mathrm{min}}$ = 0.91$\,$GeV and $M_{\mathrm{max}}$ = 1.4$\,$GeV. 
The results are shown in Fig. \ref{fig:BorelSR_0.91_1.4_mock}. 
When using $m_{\mathrm{flat}}(\omega)$, the resulting spectral function at high energy 
oscillates around the continuum value before approaching the default model.  
This is plausible, because the input spectral function has the ``pole + continuum" structure, 
which the MEM analysis is trying to simulate with the given limited information of $G_{\mathrm{mock}}(M)$. 
Nevertheless, the MEM procedure cannot reproduce the mass of the nucleon, whatever default model or 
Borel mass range is used. 

\subsection{Analysis using OPE data}
\label{subsection Analysis using OPE data}
\begin{figure}[htbp]
 \begin{center}
  \includegraphics[scale=0.68]{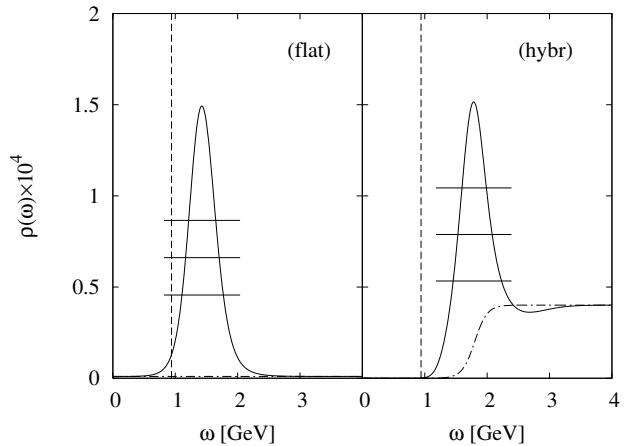}
 \end{center}
 \caption{Spectral functions extracted from $G_{\mathrm{OPE}}(M)$ using various default models at $\beta=-1$. 
The analyzed Borel mass region is 0.91 GeV $\leq M \leq$ 0.97 GeV. 
The solid line shows $\rho(\omega)$, the vertical dashed line indicates 
the position of the nucleon ground state and the 
dash-dotted line stands for the default model.}
 \label{fig:BorelSR_0.91_0.97_OPE}
\end{figure}

Similar to the previous section, we now carry out the analysis using the real OPE data, 
$G_{\mathrm{OPE}}(M)$, 
even though from our experience of the mock data analysis, we cannot expect to obtain 
meaningful results. 
We analyze the spectral function by setting 
the analyzed Borel mass region to 0.91 GeV $\leq M \leq$ 0.97 GeV, as in the last section. 
The results are shown in Fig. \ref{fig:BorelSR_0.91_0.97_OPE}. 
Using a wider Borel mass region leads to spectral functions 
very similar to the ones shown in 
Fig. \ref{fig:BorelSR_0.91_1.4_mock}. 
As in the mock data analysis, the MEM procedure does not succeed to reproduce the nucleon peak. 
The main reason for this failure can be traced back to the slow convergence of the OPE and 
to the large contribution of continuum to the sum rule. 
These factors severely reduce the information of the lowest nucleon pole that can be extracted from 
$G_{\mathrm{OPE}}(M)$.

\section{Analysis using the Gaussian sum rule}
\label{V}
In case of the Gaussian sum rule, the analysis is carried out using the following equations: 
\begin{equation}
\begin{split}
G_{\mathrm{OPE}}(s,\tau ) &= G_{\rho}(s,\tau)=\int^\infty _0d\omega K(s,\tau ,\omega)\rho(\omega), \\
K(s,\tau ,\omega) &= \frac{2\omega^{5}}{\sqrt{4\pi \tau}} 
e^{-\frac{(\omega^2-s)^2}{4\tau}}. \nonumber
\label{eq:gaussiansumrule}
\end{split}
\end{equation}
As before, we set $\beta=-1$ and the results will be shown in terms of the dimensionless spectral 
function. 

\subsection{Analysis using mock data}
\label{subsection G Analysis using mock data} 
As for the Borel sum rule, 
we use Eq.(\ref{eq:mockspec}) as mock data and Eqs.(\ref{eq:default flat} - 
\ref{eq:default hybr}) for the default model. 
As a first step, 
we must determine the ranges of s, $\tau$ used in the analysis. 
From the property of the kernel, we expect that 
$G _{\mathrm{\rho }}$ at small values of $\tau$ will be more sensitive to 
narrow structures such as the lowest peak, 
while $G _{\mathrm{\rho }}$ at larger values will to a large extent be fixed by the continuum. 
Hence, to extract as much information as possible from $G_{\mathrm{mock}} (s,\tau )$, 
we use several value of $\tau$ at the same time, which are $\tau$ = 0.5, 1, 1.5, 2$\,\mathrm{GeV}^4$. 
From 
\begin{equation}
\lim_{\tau \rightarrow 0}\frac{1}{\sqrt{4\pi \tau}} \mathrm{e}^{-\frac{(\omega ^2 -s)^2}{4\tau }}
=\delta (\omega ^{2} -s),
\end{equation}
one understands that in the limit $\tau \to 0$, $G_{\mathrm{\rho }}(s,\tau )$ 
in principle should approach the spectral function as 
\begin{equation}
\lim_{\tau \rightarrow 0} G_{\mathrm{\rho }}(s,\tau ) = s^{2} \rho (\sqrt{|s|}).
\end{equation}
However, it is not possible to take this limit, because the OPE is an expansion in powers 
of $1/\sqrt{\tau}$ as can be seen in Eq.(\ref{eq:opegaussian}), meaning that the 
convergence of the expansion worsens with lower values of $\tau$. 
\begin{figure}[tbp]
 \begin{center}
  \includegraphics[scale=0.6]{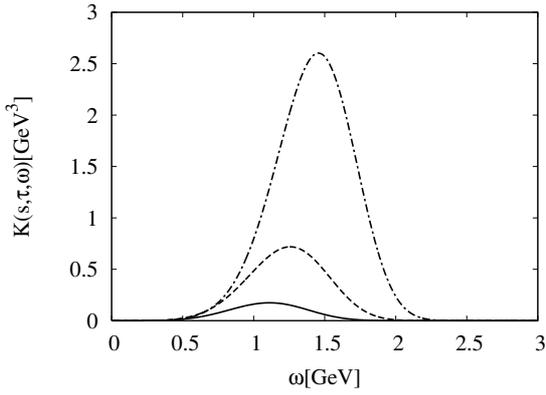}
 \end{center}
 \caption{Three examples of the kernel $K(s,\tau ,\omega)$. 
The solid line shows $K(s,\tau ,\omega)$ for $s = -0.79\,\mathrm{GeV}^2$, 
the dashed line for $s = 0\, \mathrm{GeV}^{2}$ and the dash-dotted line for $s = 0.94\,\mathrm{GeV}^2$. 
In all three cases, $\tau$ is fixed to $\tau=0.5\, \mathrm{GeV}^{4}$.}
 \label{fig:kernels}
\end{figure}
\begin{table}[t]
\begin{center}
\begin{tabular}{|c|c|c|c|c|}
\hline
$\tau\,[\mathrm{GeV}^{4}]$&0.5&1&1.5&2 \\ \hline
$s_{\mathrm{min}}\,[\mathrm{GeV}^{2}]$&-0.79&-1.84&-2.96&-4.13 \\ \hline
$s_{\mathrm{max}}\,[\mathrm{GeV}^{2}]$&0.94&0.02&-0.97&-2.0 \\ 
\hline
\end{tabular}
\caption{The analyzed ranges of the variable $s$ for the employed values of 
$\tau$.}
\label{The analyzed range of the variable s}
\end{center}
\end{table}

Turning now to the parameter $s$, similar to the Borel sum rule case, we determine 
its range from 
the convergence in $G_{\mathrm{OPE}}(s,\tau )$ 
and the contribution from the continuum in $G_{\rho }(s,\tau )$.
The analyzed regions of $s$ and $\tau$ are shown in Table \ref{The analyzed range of the variable s}. 
Here, 
we denote the upper and lower boundaries of $s$ at each $\tau$ as $s_{\mathrm{max}}$ and 
$s_{\mathrm{min}}$, 
respectively. 

Some further comments are in order here. 
Firstly, let us briefly explain the method of using the two variables of $s$ and $\tau$ at the same time, 
compared to the discussion of Section \ref{III}, where 
only one variable $x$ was considered. 
Generalizing the method from one variable to two is straightforward, 
as one only needs to redefine the kernel and the likelihood function. 
Specifically, we use 
\begin{equation}
\begin{split}
G(x_{i} ) &= \sum _{\omega _{j} = \omega _{\mathrm{min}}} ^{\omega _{\mathrm{max}}}
K(x_{i}, \omega _{j}) \rho (\omega _{j} ) 
d\omega \nonumber\\
&\downarrow  \nonumber\\
G(s_{i}, \tau ^{k}) &= \sum _{\omega _{j} = \omega _{\mathrm{min}}} ^{\omega _{\mathrm{max}}}
K(s_{i},\tau ^{k}, \omega _{j}) \rho (\omega _{j} ) 
d\omega , \nonumber
\end{split}
\end{equation}
and
\begin{equation}
\begin{split}
L(\rho )&=\frac{1}{2(x_{\mathrm{max}}-x_{\mathrm{min}})}
\sum ^{x_{\mathrm{max}}}_{x_{i} = x_{\mathrm{min}}}dx
\frac{[G_{\mathrm{OPE}}(x_{i})-G_{\rho }(x_{i})]^2}{\sigma ^{2}(x_{i})}\nonumber\\
&\downarrow  \nonumber\\
L(\rho )&= \sum ^{\tau ^{\mathrm{max}}}_{\tau ^{k}=\tau _{\mathrm{min}}}
\sum ^{s_{\mathrm{max}}^k }_{s_{i} = s_{\mathrm{min}}^k }ds^{k}
\frac{1}{2(s_{\mathrm{max}}^k-s_{\mathrm{min}}^k)}\nonumber\\
& \quad \quad \quad \quad \quad \quad \quad 
\times  \frac{[G_{\mathrm{OPE}}(s_{i},\tau ^{k})-G_{\rho }(s_{i},\tau ^{k})]^2}
{\sigma ^{2}(s_{i},\tau ^{k})}.\nonumber
\end{split}
\end{equation}
In the present analysis, we take 4 data points for $\tau$ ($N_{\tau }=4$), 
and 25 data points for s ($N_s=25$) at each $\tau ^{k}$ and
adjust $\omega _{\mathrm{min}}$ and $\omega _{\mathrm{max}}$ to 0 and 6 GeV, respectively. 
$s_{\mathrm{max}}^{k}$ and $s_{\mathrm{min}}^{k}$ at each $\tau ^{k}$ are given in 
Table \ref{The analyzed range of the variable s}, 
and $ds^{k}$ is $(s_{\mathrm{max}}^k-s_{\mathrm{min}}^k)/N_{s}$.
\begin{figure*}[!tbp]
 \begin{center}
  \includegraphics[clip,scale=1.0]{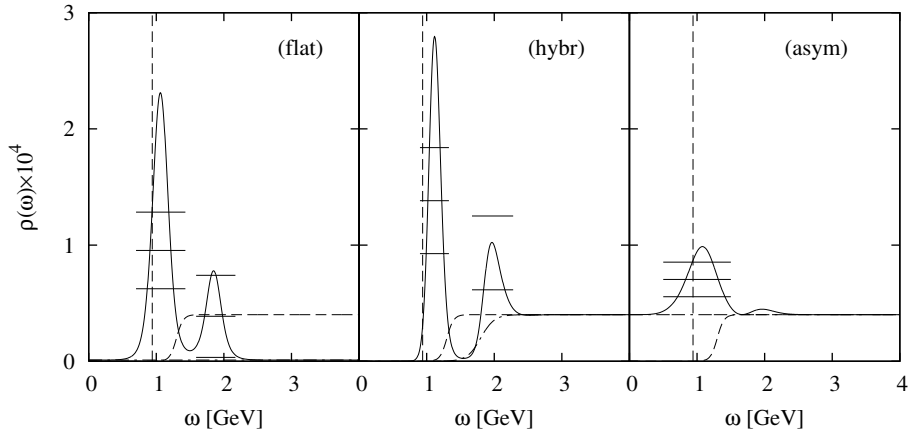}
 \end{center}
 \caption{Spectral functions extracted from $G_{\mathrm{mock}}(s,\tau)$ using various default models 
at $\beta=-1$. 
The analyzed regions of $s$ and $\tau$ are shown in Table \ref{The analyzed range of the variable s}. 
The solid line shows $\rho(\omega)$, the dashed line depicts the input mock spectral function and 
the dash-dotted line stands for the default model.}
 \label{fig:GuassianSR_smax_cont_mock_beta=-1}
\end{figure*}
\begin{table*}[htbp]
\begin{center}
\begin{tabular}{|c|c|c|c|c|}
\hline
 &mock data&$m_{\mathrm{flat}}$&$m_{\mathrm{hybr}}$&$m_{\mathrm{asym}}$\\ \hline
position of first peak [MeV]&940&1060 $\pm$ 140&1120 $\pm$ 100&1080 $\pm$ 410 \\ \hline
position of second peak [MeV]&-&1850&1970&1960 \\ \hline
residue of first peak $[10^{-5}\mathrm{GeV}]$&8.3&7.0&6.0&- \\
\hline
\end{tabular}
\caption{Position of the peaks and values of the corresponding 
residues, obtained from a mock data analysis 
with three default models at $\beta=-1$. 
Here, the given values of the mock data residue is defined as 
$\lambda^{2}/(2M_{N}^{5})$. 
The errors are estimated by evaluating 
the half width of the obtained first peak.}
\label{table3}
\end{center}
\end{table*} 

Secondly, we comment on the ranges of the variable $s$, shown in Table \ref{The analyzed range of the variable s}. 
It is seen that to the most part, we use negative values for $s$, which may seem somewhat counterintuitive as, 
naively, the kernel of Eq.(\ref{eq:gaussiansumrule}) seems to have a peak at $\sqrt{s}$, when considered as a function 
of $\omega$. Therefore one would expect values of $s$ around 1 GeV to give a $G_{\rho}(s,\tau)$ that is most 
sensitive to the spectral function in the region of the nucleon pole. However, this is not the case because 
the kernel is distorted due to the $\omega^5$-factor in front of the Gaussian of Eq.(\ref{eq:gaussiansumrule}). 
For example, for constructing a kernel with a maximum value at 
1 GeV which is the energy region which we are mostly interested in, 
$s$ becomes $1-5\tau $ and, in the case of $\tau = 0.5, 1, 1.5, 2$ GeV$^{4}$, 
has a negative value. 
This is illustrated in Fig. \ref{fig:kernels}, where one can see that only kernels with negative $s$ have 
peaks around 1 GeV. From these arguments, we can understand that it is necessary and important to use negative 
values of $s$ in the analysis of the Gaussian sum rules. 

Let us now discuss the 
obtained spectral functions, which are shown in Fig. \ref{fig:GuassianSR_smax_cont_mock_beta=-1}. 
Their detailed numerical results are given in Table \ref{table3}. 
It is observed that, 
compared with the Borel sum rule, the reconstruction of the lowest lying peak 
has considerably improved. 
Nevertheless, it is seen in Table \ref{table3} that its position 
is shifted upwards about $120 \sim 180\,$MeV, which should give an idea 
of the precision attainable with this method.
To evaluate the residue of the first peak, we first have to define the region $\omega$ over 
which the peak can be considered to be the dominant contribution. 
The left edge of the first peak is determined to be the point whose height is 
1/30 of the peak vertex 
and the right edge so that the center between the left and the 
right edge lies on the peak vertex. 
The residue is then obtained by simply integrating $\rho (\omega )$ over the peak region, 
indicated by the horizontal length of the error bars. 
Table \ref{table3} shows that 
the residue of the first peak gives about 80 \% 
of the input value of the mock data. 
Since the default model $m_{\mathrm{flat}}(\omega)$ leads to a peak position and 
residue closest to the input values we will use only $m_{\mathrm{flat}} (\omega )$ 
as the default model in the following. 
 
Here, we comment on the width of the peaks appearing in the obtained spectral functions. 
One might wonder why these peaks have a finite width even though the input mock data only 
contains a zero width pole. There are two important aspects related to this point. 
Firstly, one should note that the OPE side of the sum rule is rather insensitive to the value 
of the width, because this information is largely lost after the integration over the spectral 
function in Eqs. (\ref{eq:BorelSR}) or (\ref{eq:GaussianSR}). 
This is so as long as the parameters governing the scales of these integrals ($M$ or $\tau^{1/4}$) 
are much larger than the width of the peak of interest. 
This point was discussed explicitly 
for the $\rho$-meson channel in \cite{Gubler1}. 
Secondly, one should remember that MEM is a statistical method, that can only provide the 
the most probable form of the spectral function, given the information available. 
This most likely spectral function depends 
not only on the input data but also on their error. 
Generally, a larger error makes a peak broader and smaller in magnitude, 
therefore introducing an artificial width connected to the error of the OPE data.

\begin{figure*}[!tbp]
 \begin{center}
  \includegraphics[scale=1.0]{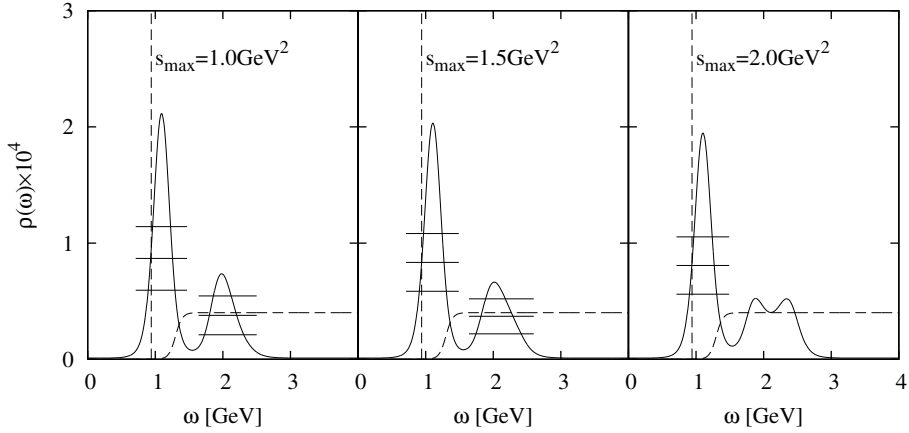}
 \end{center}
 \caption{Spectral functions extracted from $G_{\mathrm{mock}}(s,\tau)$ and $m_{\mathrm{flat}}(\omega)$ at $\beta=-1$. 
Various values of $s_{\mathrm{max}}$ are used as indicated in the figures. 
The values of $s_{\mathrm{min}}$ at different $\tau$ are shown in Table \ref{The analyzed range of the variable s}. 
The solid line shows $\rho(\omega)$ and the dashed line stands for the input mock spectral function.}
 \label{fig:GuassianSR2}
\end{figure*}

Next, to study the dependence of the results on the range of $s$, we fix $s_{\mathrm{max}}$ to 
1.0$\,\mathrm{GeV}^2$, 1.5$\,\mathrm{GeV}^2$ and 2.0$\,\mathrm{GeV}^2$ for all values of $\tau$, and redo 
the analysis. 
The results are shown in Fig. \ref{fig:GuassianSR2} and Table \ref{table4}. 
From Table \ref{table4}, one can observe that 
the position and residue of the first peak depends on $s_{\mathrm{max}}$ only weakly, while 
the position of the second peak is quite sensitive to the value of $s_{\mathrm{max}}$. 
Furthermore, from the Fig. \ref{fig:GuassianSR2}, it can be understood that 
the second peak rather 
represents the continuum, around which the MEM output oscillates. 
\begin{table*}[!tbp]
\begin{center}
\begin{tabular}{|c|c|c|c|}
\hline
$s_{\mathrm{max}}\,[\mathrm{GeV}^{2}]$&$s_{\mathrm{max}}=1.0$
&$s_{\mathrm{max}}=1.5$&$s_{\mathrm{max}}=2.0$\\ \hline
position of first peak [MeV]&1090 $\pm$ 140&1100 $\pm$ 140&1100 $\pm$ 150\\ \hline
position of second peak [MeV]&1980&2020&- \\ \hline
residue of first peak $[10^{-5}\mathrm{GeV}]$&6.7&6.5&6.3 \\
\hline
\end{tabular}
\caption{Position of the peaks and values of the corresponding residues. 
Here, the default model $m_{\mathrm{flat}}(\omega)$ and various $s_{\mathrm{max}}$ 
values were used. 
The errors are estimated by evaluating 
the half width of the obtained first peak.}
\label{table4}
\end{center}
\end{table*}

\begin{figure}[tbp]
 \begin{center}
  \includegraphics[scale=0.68]{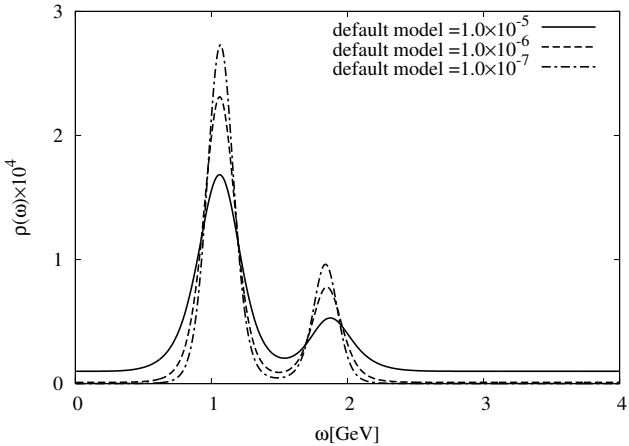}
 \end{center}
 \caption{
The spectral functions ectracted from $G_{\mathrm{mock}}(s,\tau)$, using various height values for 
the flat default model $m_{\mathrm{flat}}(\omega)$. 
The concrete numerical values are given in the figure. 
}
 \label{fig:comparingflat}
\end{figure}

Finally, we discuss the dependence of the spectral function on 
the parameters appearing in the default models. 
In the case of the flat default models as given in Eqs.(\ref{eq:default flat}) and (\ref{eq:default asym}), we
have carried out analyses 
using several height values (spanning over several orders of magnitude).
The results are shown in Fig. \ref{fig:comparingflat}. 
It is clear that the detailed values
 have little influence on the position of the nucleon pole. 
Extracting the correspoding residues, it is understood that their values 
also depend on the height value only weakly. 
Specifically, the pole positions 
for the curves shown in Fig. \ref{fig:comparingflat} 
coincide within 1\% and the residues within 6\%. 
In the case of a hybrid default model as given in Eq.(\ref{eq:default hybr}), 
we also carried out analyses using several parameter combinations,
the result being that
 the positions of the obtained first peaks did depend on the
parameter combinations only weakly.

In all, we conclude that using the Gaussian sum rule, 
reconstruction of the lowest lying peak and the residue of mock data is largely improved 
compared to the Borel sum rule, and that the attainable precision with the 
present uncertainties of the condensates is about 20 \%. 

\subsection{Analysis using OPE data}
\label{subsection G Analysis using OPE data}
\begin{figure*}[!tbp]
 \begin{center}
  \includegraphics[scale=0.9]{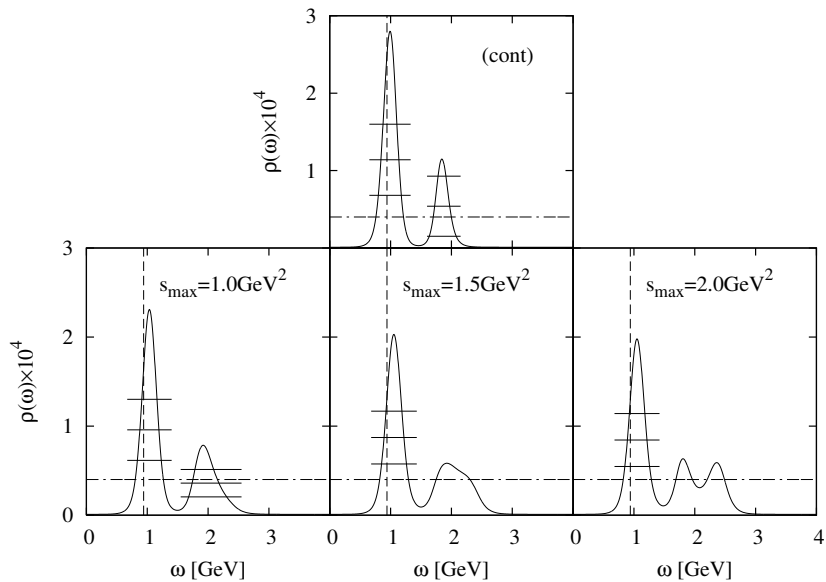}
 \end{center}
 \caption{Spectral functions extracted from the MEM analysis using $G_{\mathrm{OPE}}(s,\tau)$ 
and $m_{\mathrm{flat}}(\omega)$ at $\beta=-1$. 
The solid lines show $\rho(\omega)$ and 
the dash-dotted lines stands for the asymptotic value of the spectral function at high energy. 
(Note that $\rho(\omega )$ is normalized by $\frac{1}{\omega ^{4}}$.) 
In the top figure (cont), the values of $s$ and $\tau$ given in 
Table \ref{The analyzed range of the variable s} are used, 
while for the other figures the values of $s_{\mathrm{max}}$ are fixed as indicated.} 
\label{fig:GuassianSR3}
\end{figure*}

Next, we carry out the analysis using $G_{\mathrm{OPE}}(s, \tau )$. 
We apply the MEM in the same regions of $s$ and $\tau$ used in the mock data analysis. 
The results are shown in Fig. \ref{fig:GuassianSR3}, while the numerical details are given in 
Table \ref{table5}. 
The behavior of the results is similar to those of the mock data analysis. 
We observe that the positions of the lowest peak 
lie quite close to the experimental value. 
Besides a clearly resolved first peak, the spectral function exhibits one or two higher peaks which 
oscillate around the 
asymptotic high energy limit. In principle, this part could also contain nucleon resonances with both positive and 
negative parity. With the present resolution achievable with the MEM technique, it however seems to be rather difficult 
to resolve these resonances from the continuum. Using the parity projected sum rules \cite{Jido}, which 
can also be analyzed by the MEM approach, might improve the situation. 

\subsection{Investigation of the $\beta$ dependence}
\label{subsection V3}
\begin{figure*}[!tbp]
 \begin{center}
  \includegraphics[scale=1.1]{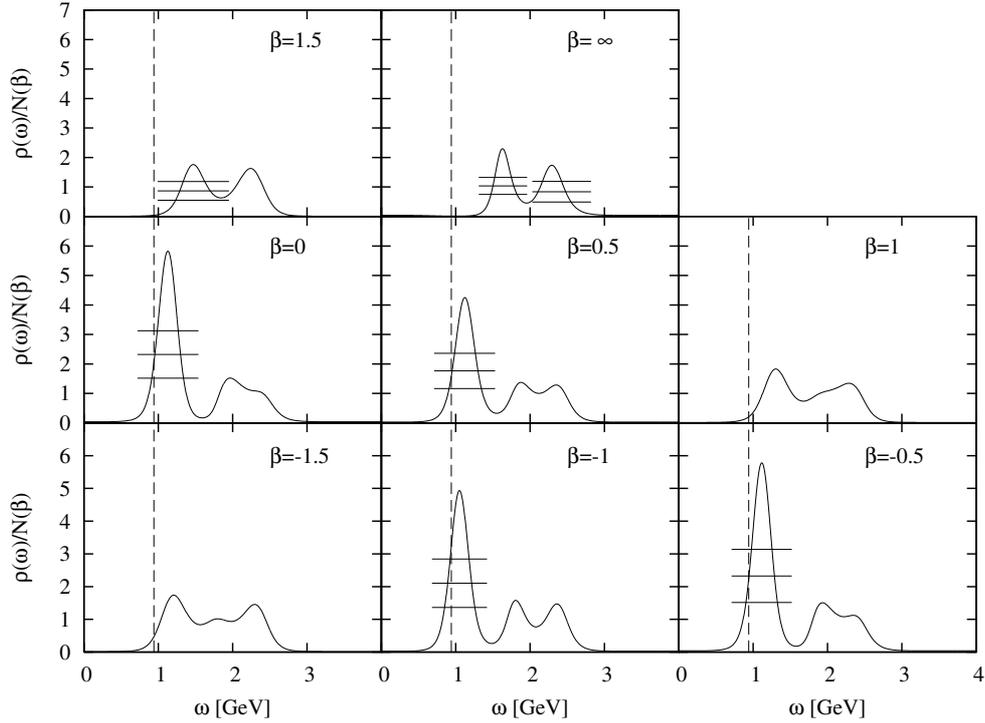}
 \end{center}
 \caption{Spectral functions 
obtained from the MEM analyses of OPE data  
for various values of $\beta$, normalized by the factor $N(\beta)$.
The values $s_{\mathrm{min}}$ are given in Table \ref{table6} and $s_{\mathrm{max}}$ 
is fixed to 2.0 GeV$^2$.}
 \label{fig:G5}
\end{figure*}

The coupling strengths of the nucleon ground state 
and its excited states depend on the choice of the interpolating operator, 
\textit{i.e.}, on the value of $\beta$.
To investigate the nature of this dependence, 
we let $\beta$ vary as $-$1.5, $-$0.5, 0.0, 0.5, 1.0, 1.5, $\infty$ and 
extract the corresponding spectral functions. For $\tau$, we have chosen 
0.5, 1.0, 1.5, 2.0$\,\mathrm{GeV}^4$ as before and $s_{\mathrm{min}}$ 
is determined from the OPE convergence condition at each $\beta$. 
In the case of $\beta=1$, $s_{\mathrm{min}}$ is determined from the dimension 4 term 
because the higher dimensional terms vanish for this choice.
To obtain information not only on the spectral function around 1 GeV, but also in 
the region of possible excited states, $s_{\mathrm{max}}$ is not determined 
from the dominance of the lowest lying state in $G_{\rho}(s,\tau)$ 
but is fixed to 2.0 GeV$^{2}$. 
The explicit values of $s_{\mathrm{min}}$ are given in Table \ref{table6}. 
Note that $\beta = \infty$ should not be taken literally, but just means that 
we use the correlator of only $\eta_2(x)$ (which is the 
coefficient of $\beta^2$ in Eq.(\ref{eq:ope})) for the analysis. 
The resulting spectral functions and their numerical properties are shown in 
Fig. \ref{fig:G5} and Table \ref{table7}. 
For a better comparison, we have normalized the spectral functions in Fig. \ref{fig:G5} 
by dividing them by the factor $N(\beta) \equiv (5 + 2\beta + 5\beta^2)/(128(2\pi)^4)$, 
so that the continuum approaches unity in the high energy limit. 
\begin{table*}[!htbp]
\begin{center}
\begin{tabular}{|c|c|c|c|c|c|}
\hline
$s_{\mathrm{max}}\,[\mathrm{GeV}^{2}]$&$s_{\mathrm{max}}$=(cont)&$s_{\mathrm{max}}$=1.0&$s_{\mathrm{max}}$=1.5&
$s_{\mathrm{max}}$=2.0 \\ \hline
position of first peak$\,$[MeV]&990 $\pm$ 130&1040 $\pm$ 140&1060 $\pm$ 150&1050 $\pm$ 150 \\ \hline
position of second peak$\,$[MeV]&1840&1920&-&1800 \\ \hline
residue of first peak$\,[10^{-5}\mathrm{GeV}]$&7.9&7.0&6.6&6.3 \\ %\hline
\hline
\end{tabular}
\caption{Positions of the first and second peaks and the corresponding residues 
of the first peak obtained from the analyses using OPE data at $\beta=-1$
and the default model $m_{\mathrm{flat}}(\omega)$. The errors are estimated by evaluating 
the half width of the obtained first peak.}
\label{table5}
\end{center}
\end{table*}
\begin{table*}[!t]
\begin{center}
\begin{tabular}{|c|c|c|c|c|c|c|c|c|}
\hline 
$\beta$      &-1.5 &-1&-0.5 &0    &0.5  &1    &1.5  &$\infty $    \\ \hline
$\tau=$0.5   &-0.40&-0.79&-0.39&-0.35&-0.43&0.92 &-0.23&1.04  \\ \hline
$\tau=$1.0   &-1.21&-1.84&-0.96&-0.87&-1.10&0.12 &-0.93&-0.36 \\ \hline
$\tau=$1.5   &-2.12&-2.96&-1.63&-1.47&-1.87&-0.76&-1.74&-0.88 \\ \hline
$\tau=$2.0   &-3.07&-4.13&-2.34&-2.13&-2.68&-1.66&-2.60&-1.45 \\ 
\hline
\end{tabular}
\caption{Values of $s_{\mathrm{min}}$ [GeV$^2$] at each $\beta$ and $\tau$ [GeV$^4$]. 
$s_{\mathrm{max}}$ is fixed to 2.0 GeV$^2$.}
\label{table6}
\end{center}
\end{table*}

\begin{figure*}[tbp]
 \begin{center}
  \includegraphics[scale=1.0]{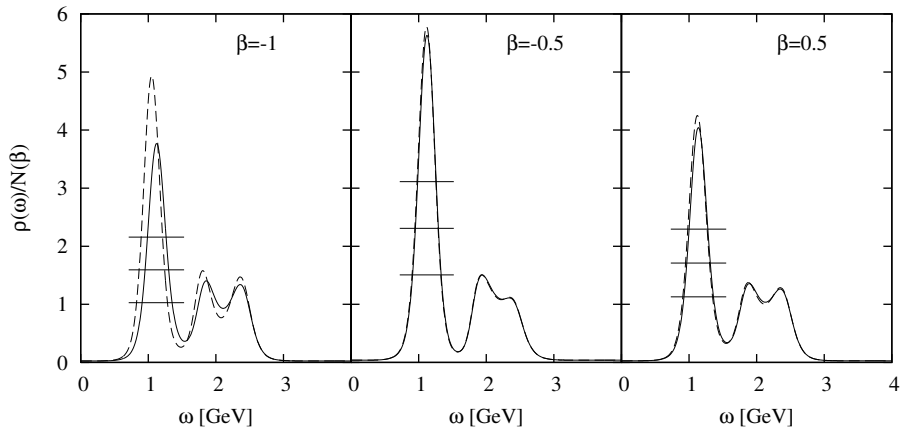}
 \end{center}
 \caption{
Spectral functions obtained from a MEM analysis of OPE data with 
$m_{\mathrm{flat}}(\omega)$ at $\beta= -1$, $-0.5$ and $0.5$, normalized by $N(\beta)$. 
The solid lines show the normalized output spectral function $\rho (\omega )$, 
for which a $\beta$ independent ``refined" criterion was used to determine $s_{\mathrm{min}}$. 
The details are explained in the text. 
For comparison, the spectral functions obtained with the previous $\beta$-dependent 
criterion are indicated by the dashed line. 
}
 \label{fig:G77}
\end{figure*}

Considering first the lowest peaks at $\omega \sim $ 1 GeV, we see that a peak is 
clearly resolved for $\beta$ = $-$1.0, $-$0.5, 0.0 and 0.5, located at 
about the same position and with similar residue values. For the other values of $\beta$, 
no prominent peak is observed. 
These results are, however, obtained by using the $s_{\mathrm{min}}$ values determined separately 
for each $\beta$ 
and one could suspect that 
the choice of $s_{\mathrm{min}}$ affects the properties of the first peak. 
To study this dependence and to get an idea of the stability of our results, 
we redo the analysis, now using values of $s_{\mathrm{min}}$ that are 
determined via a $\beta$ independent criterion as advocated in \cite{Leinweber2,Leinweber3}. 
Taking a closer look at Table \ref{table6}, it is observed that in the region between 
$\beta = -1.0$ and $0.5$, where the ground state peak can be extracted, the OPE convergence is 
worst for $\beta = 0.0$, giving the largest value of $s_{\mathrm{min}}$. This implies that 
it is most reasonable to fix $s_{\mathrm{min}}$ to the values of $\beta=0.0$, so that 
no values of $s$ are used, where the convergence of the OPE might be questionable. 
Using this criterion, 
we repeat the analysis at $\beta$ = $-$1, $-$0.5, 0.5. As before, we set $s_{\mathrm{max}}$ to 
2.0 GeV$^2$.
\begin{table*}[!t]
\begin{center}
\begin{tabular}{|c|c|c|c|c|c|c|c|c|}
\hline
$\beta$                         &-1.5&-1  &-0.5&0.0  &0.5  &1   &1.5 &$\infty$    \\ \hline
position of first peak$\,$[MeV]    &1210&1050&1120&1130&1120&1300&1470&1630 \\ \hline
residue of first peak$\,[10^{-5}\mathrm{GeV}]$ &-   & 6.3&5.0 &4.8 &5.3 &-   &-   &-    \\ 
%$\tau[GeV^{4}]$=1.5&-2.12&-1.63&-1.47&-1.87&-3.55&-0.76&-1.74&-0.90 \\ \hline
%$\tau[GeV^{4}]$=2  &-3.07&-2.34&-2.13&-2.68&-4.92&-1.66&-2.60&-1.49 \\ 
\hline
\end{tabular}
\caption{Positions of the first peaks of the spectral functions shown in Fig. \ref{fig:G5}. 
For the cases in which the first peak can be interpreted as the nucleon ground state, 
the corresponding residue is given as well.}
\label{table7}
\end{center}
\end{table*} 
The results are shown and compared to the previous ones in Fig. \ref{fig:G77}. The 
numerical details are given in Table \ref{table98}.  
We see from these results that some details of the output spectral functions (especially at 
$\beta=-1$) change when the new criterion is used. The qualitative structure of a clear 
lowest peak with a continuum structure at a somewhat higher energy is however not altered. 

\begin{figure}[tbp]
 \begin{center}
  \includegraphics[scale=0.68]{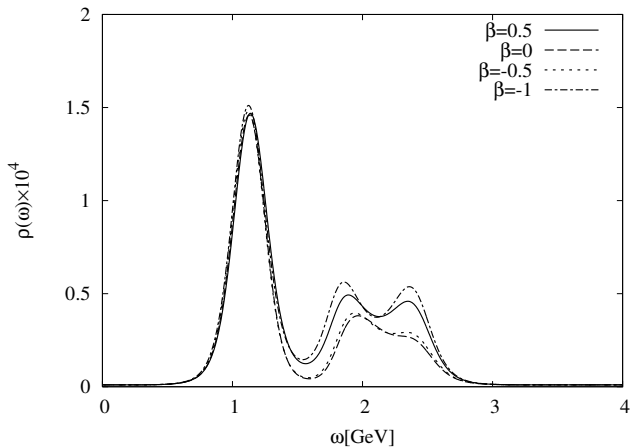}
 \end{center}
 \caption{
Non-normalized spectral functions obtained from a MEM analysis using OPE data and 
$m_{\mathrm{flat}}(\omega)$ at $\beta$ = $-$1.0, $-$0.5, 0, 0.5. 
The employed $s_{\mathrm{min}}$ is fixed to the one at $\beta=0$ and 
$s_{\mathrm{max}}$ to 2.0$\,$GeV$^2$.}
 \label{fig:comparingrho1}
\end{figure}

We also observe from Table \ref{table98} that the position and residue 
are almost independent of $\beta$. 
This fact becomes even more explicit when we plot the non-normalized spectral functions 
for several values of $\beta$ around $\beta=0$ as shown in Fig. \ref{fig:comparingrho1}. 
This figure clearly illustrates that, in contrast to the continuum spectra, the property 
of the lowest peak is essentially $\beta$-independent. 

All these results can be naturally interpreted by presuming that the 
lowest lying pole couples strongly to $\eta _1$, and only very weakly to $\eta _2$. 
Therefore, the pole appears to be almost independent of $\beta$ as long as $|\beta |$ is small, but disappears 
when $|\beta |$ becomes large and the contribution of $\eta _2$ dominates the spectral function. 
The nucleon ground state pole can be resolved as long as its strength is large enough 
compared to the continuum contribution. At about $|\beta| \geq 1.5$, however, this 
continuum contribution gets too large to extract 
information on the lowest peak. These conclusions agree with the findings of 
lattice QCD \cite{Leinweber3,SSasaki,Melnitchouk} and some earlier suggestions of 
QCD sum rule studies \cite{Chung2}, in which however the position of the nucleon 
ground state peak was used as an input. 

Furthermore, by looking at Table \ref{table6} and Fig. \ref{fig:G77}, we can also understand 
why it is the Ioffe current ($\beta=-1.0$), rather than $\eta_1$, that seems to be 
most suitable for QCD sum rule studies. This comes from the fact that the OPE 
convergence is considerably faster for the Ioffe compared to $\eta_1$, which allows the 
analysis to become more sensitive to the lower energy region of the spectral function, 
therefore providing more information on the nucleon peak.
\begin{table}[t]
\begin{center}
\begin{tabular}{|c|c|c|c|c|c|}
\hline
$\beta$                &-1   &-0.5&0&0.5    \\ \hline
position of first peak$\,$[MeV]  &1120 &1120&1130 &1140\\ \hline
residue of first peak$\,[10^{-5}\mathrm{GeV}]$&5.3  &4.9 &4.8&5.1 \\ 
\hline
\end{tabular}
\caption{Positions of the first peaks and the corresponding residues 
of the spectral functions shown in Fig. \ref{fig:G77}, where a $\beta$-independent 
criterion was used to fix $s_{\mathrm{min}}$. For comparison, the result for $\beta$=0 is shown as well.}
\label{table98}
\end{center}
\end{table}

As a last point, let us also consider the continuous structure above the nucleon ground state pole and 
possible excited states appearing there. It can be observed in Fig. \ref{fig:G5} that for most 
values of $\beta$, the spectral function just oscillates around the high energy limit, similar 
to the results obtained from the mock data analysis shown in Fig. 
\ref{fig:GuassianSR2}, where we have only included the continuum and 
no resonances into the input spectral function. On the other hand, for $\beta = 1.5$ and 
especially $\beta=\infty$, the behavior is somewhat different, showing a quite clear first 
peak at about 1.6$\,$GeV. This peak could correspond to one of the nucleon resonances, 
N(1440), N(1535), N(1650) or a combination of them. 
Results of lattice QCD studies \cite{SSasaki,Melnitchouk} indicate that
$\eta_2$ couples 
to the negative parity states N(1535) and N(1650), while the 
first excited positive parity state lies considerably above the Roper 
resonance N(1440). 
Even though the systematic uncertainties of these calculations are probably not 
yet completely under control, 
these findings suggest that
the lowest peak seen at $\beta = \infty$ corresponds rather to N(1535)
or N(1650)
and not to N(1440).
It, however, seems to be difficult at the 
present stage to make any conclusive statements on the nature of this peak. 
An analysis of the parity projected sum rules will hopefully help to clarify this 
issue. 

\section{Summary and Conclusion}
\label{VI}
In this study, we have applied the MEM technique, which is based on Bayes' theorem of probability theory 
to the analysis of the nucleon QCD sum rules. We have investigated two kinds of sum rules, 
namely the frequently used Borel sum rule and the less known Gaussian sum rule. 
Before analyzing the actual sum rules, we have first tested the applicability of the MEM 
approach by constructing and analyzing realistic mock data. Our findings show that due to 
the properties of the kernel and the slow OPE convergence, it appears to be difficult to 
extract much meaningful information on the nucleon ground state from the Borel sum rule. 
Another reason for this failure may also be that, 
because only spectral functions satisfying positivity can be analyzed with the currently 
available MEM procedure, 
we cannot use the chiral odd part 
$\Pi_2(q^2)$ of Eq.(\ref{eq:correlation function0}) 
in the present analysis, which has been claimed to be 
more reliable. 
For instance, in \cite{Leinweber2} 
analyses using only
$\Pi_1(q^2)$ or $\Pi_2(q^2)$ were carried out and the respective Borel windows 
examined. As a result,
in case of only using
$\Pi_1(q^2)$, the Borel window is seen to be very narrow, 
making it difficult to obtain a reliable estimate for the 
nucleon mass. 
On the other hand, in case of using
only $\Pi_2(q^2)$, the borel window was shown to be sufficiently large,
so that the nucleon mass can be reliably obtained. 
It therefore would be helpful if $\Pi_2(q^2)$ could be used. 
As long as one uses MEM, this, 
however, will only become possible when the parity projected 
sum rules are employed. 
As an alternative, we have formulated the Gaussian sum rule and found 
that it allows us to extract more information 
on the spectral function and enables us to reconstruct the nucleon ground 
state with reasonable precision from both the mock and the OPE data. 

As the analysis is done with the MEM technique, we obtain the spectral function directly and 
do not have to deal with quantities depending on unphysical parameters such as the Borel mass. 
Moreover, we do not have to restrict the spectral function to the traditional ``pole + 
continuum" form, which allows us to investigate the spectral function of a large variety of 
of interpolating fields. From this investigation, we have found that the nucleon pole is independent 
of the parameter $\beta$ of Eq.(\ref{eq:interpolating field}) and vanishes when $\eta_2$ 
becomes the dominant contribution of the correlator. Thus we conclude that, 
in agreement with findings of lattice QCD, the nucleon ground state couples only to the interpolating 
field $\eta_1$, but not (or only very weakly) to $\eta_2$. Furthermore, a 
peak structure is seen around 1.6 GeV in the spectral function corresponding to $\eta_2$, which 
suggests that some nucleon resonance in this region couples to $\eta_2$. To clarify the 
nature of this peak, more thorough investigations are needed. 

In all, we have shown that the MEM technique in combination with the Gaussian sum rule formulated 
in this paper is useful for extracting the properties of the nucleon ground state and may even 
make it possible to investigate possible excited states. 
There are, however, still several open 
questions to be answered. First of all, we have so far ignored all radiative $\alpha_s$ 
corrections to the Wilson coefficients. These corrections are known to be significant and it 
is therefore important to include them for a more quantitative analysis. Additionally, 
possible violations of the vacuum saturation approximation should also be considered. 
As a further point, it would be crucial to separate the contributions from positive and 
negative parity states to the spectral function, especially to investigate the excited nucleon resonances. 
These issues are left for future investigations.

\begin{acknowledgement}
We would like to thank Giuseppina Orlandini for her suggestion to use the Gaussian 
sum rule in combination with the Bayesian approach, which initiated parts of this study. 
This work is partially supported by KAKENHI under Contract Nos. 19540275  and 22105503. 
P.G. gratefully acknowledges the support by the Japan Society for the 
Promotion of Science for Young Scientists (Contract No. 21.8079). 
The numerical calculations of this study have been partly performed on the super grid 
computing system TSUBAME at Tokyo Institute of Technology.
\end{acknowledgement}

%
% BibTeX users please use
% \bibliographystyle{}
% \bibliography{}

\end{document}